\documentstyle[11pt]{article}
\newlength{\bredde}
\def\slash#1{\settowidth{\bredde}{$#1$}\ifmmode\,\raisebox{.15ex}{/}
\hspace*{-\bredde} #1\else$\,\raisebox{.15ex}{/}\hspace*{-\bredde} #1$\fi}
\newcommand{\beq}{\begin{equation}}
\newcommand{\eeq}{\end{equation}}
\newcommand{\bea}{\begin{eqnarray}}
\newcommand{\eea}{\end{eqnarray}}
\newcommand{\rf}[1]{(\ref{#1})}

\def\gtwid{\raise.3ex\hbox{$>$\kern-.75em\lower1ex\hbox{$\sim$}}}
\def\ltwid{\raise.3ex\hbox{$<$\kern-.75em\lower1ex\hbox{$\sim$}}}
\begin{document}
\topmargin -0.8cm
\oddsidemargin -0.8cm
\evensidemargin -0.8cm
\headheight 0pt
\headsep 0pt
\topskip 9mm
\vspace{1.5cm}
\begin{center}
{\Large{Deconfinement, Screening and Abelian Projection at Finite
Temperature}}
\footnote{Talk presented by P.H. Damgaard
at the Workshop on Non-Perturbative Approaches to QCD in Trento, Italy,
July 10-29, 1995. To appear in those proceedings (D. Diakonov, Ed.).}

\vspace{0.5cm}

{\sc Poul H. Damgaard}
\vspace{0.1cm}
\\{\sl The Niels Bohr Institute}\\
{\sl Blegdamsvej 17}\\{\sl DK-2100 Copenhagen, Denmark}
\vspace{0.4cm}
\\{\sc Jeff Greensite}
\vspace{0.1cm}
\\{\sl Dept. of Physics and Astronomy \\
       San Francisco State University \\
       San Francisco, CA 94132 USA}
\vspace{0.3cm}
\\and 
\vspace{0.3cm}
\\{\sc Martin Hasenbusch}
\vspace{0.1cm}
\\{\sl DAMTP}\\{\sl Silver Street} 
\\{\sl Cambridge CB3 9EW, England}
\end{center}
\vfill
\begin{abstract}The behavior of static sources transforming according to
different irreducible representations of the gauge group is studied in
the context of finite temperature lattice gauge theory. We combine
analytical and numerical approaches to extract information about
confinement and screening both at low temperatures, and around
the deconfinement phase transition. The idea that abelian projection in
the Maximally Abelian Gauge reproduces the most important features of
confinement and screening is tested at a quantitative level in these 
finite-$T$ theories. Our results show that while this abelian projection
provides correct qualitative features, it fails at a detailed
quantitative level.
\end{abstract}
\vfill
\begin{flushleft}
NBI-HE-95-35 \\
hep-lat/9511007
\end{flushleft}
\newpage

\section{Introduction}When one discusses the confinement properties of
pure gauge theories, one normally restricts the attention to the static
potential between two infinitely heavy sources transforming according
to the fundamental representation of the gauge group. For the purposes
of understanding the dynamical mechanisms of confinement and screening
it is of 
interest to enlarge this framework by considering static sources transforming 
according to arbitrary irreducible representations of the gauge group. 
It is also of 
interest to study an extreme situation where the system is in equilibrium
with a temperature $T$ so high that the confining properties are lost.

In this talk we shall describe some lattice gauge theory simulations that 
concern this combined situation: the behavior of static ``quark'' sources
in different representations of the gauge group, with the system being
in equilibrium with a heat bath of temperature $T$. We shall restrict
our attention to the gauge groups $SU(2)$ and $SU(3)$. Some of the results
for $SU(2)$ have been published already \cite{DH}, but the bulk of the
present contribution is based on previously unpublished material.  

The static potential
between two infinitely heavy sources is believed to depend crucially
on the manner in which the chosen representation behaves under
transformations restricted to the center $Z(N)$ of the gauge group.
Representations that are insensitive to $Z(N)$ transformations
should yield a screened potential, while those sensitive to these
transformations should yield a confining potential. This is the standard
picture of confinement and screening in non-Abelian gauge theories, 
dating back almost twenty years (see, $e.g.$, ref. \cite{Mack}).
At this conference, a major theme was one explicit picture of confinement
that {\em a priori} does not seem to refer to this $Z(N)$ symmetry
at all, that of a ``dual Meissner effect''. An explicit realization
of this picture might well depend on a suitable gauge \cite{tHooft}. The
so-called ``Maximally Abelian Gauge'' \cite{Kronfeld} has been
argued (see, $e.g.$, ref. \cite{Suzuki1}) to be superior in the sense
that it should allow one to reconstruct, for example,
the full $SU(N)$ string tension from Abelian observables --
the Abelian projections. This is a very strong statement, reducing
essentially all the dynamics of the confinement mechanism to an
Abelian dual Meissner effect in that gauge. And the r\^{o}le played
by the center symmetry is completely obscured. 

As is well known, it is a global $Z(N)$
counterpart at finite-$T$ which makes the Polyakov line an order
parameter for the confinement/deconfinement phase transition in a
pure $SU(N)$ gauge theory. With
very mild assumptions, this makes continuous deconfinement phase
transitions in $(d+1)$-dimensional $SU(N)$ gauge theories fall in the
universality classes of globally $Z(N)$-symmetric spin systems in
$d$ dimensions \cite{Svetitsky}. This promotes the global $Z(N)$ symmetry
at finite $T$ to an observable: by measuring, $e.g.$, critical
exponents one can infer which global $Z(N)$ symmetry group is being
spontaneously broken at the transition. 

Under the global $Z(N)$ transformation, the Polyakov line in the
fundamental representation transforms
as $\langle Tr_FW(x) \rangle \to z Tr_FW(x)$. The Abelian projection
of the fundamental-representation Polyakov line clearly transforms
as well, and hence serves as an order parameter on equal footing with
the full Polyakov line. But although both are order parameters for
the confinement/deconfinement phase transition, there is of course
no guarantee that they will agree even roughly in numerical values.
We shall return to this question towards the end of this talk.

\section{Finite-temperature Deconfinement: The Mean-Field Limit}

The continuous finite-$T$ deconfinement phase transitions in pure
gauge theories are particularly interesting from a theoretical point
of view because they allow us to test universality ideas in a highly
non-trivial setting. The universality predictions of Svetitsky and
Yaffe \cite{Svetitsky} give {\em exact} statements about these
transitions. It is not often we are in a position to test exact 
predictions for $SU(N)$ gauge theories, and it obviously merits a lot
of attention.

The restriction to continuous phase transitions puts some limits on
where the comparison can be carried out. But it is commonly accepted that
the (3+1)-dimensional $SU(2)$ theory has a continuous transition,
which hence should belong to the $Z(2)$ fixed point in 3 dimensions:
the 3-$d$ Ising fixed point. A lot of computational effort has gone
into verifying this by treating the Polyakov line in the fundamental
representation of $SU(2)$ as the analogue of the spin variable in
the corresponding 3-dimensional Ising model. Obvious quantities to
compare include the critical exponents, amplitude ratios etc. 
The agreement is impressive
(see, $e.g.$, ref. \cite{Bielefeld} for a recent study, and references
therein). In (3+1)-dimensions this unfortunately exhausts our chances
for testing the universality conjecture on $SU(N)$ theories, since the
deconfinement phase transitions are believed to be 1st order for all
higher values of
$N$ in (3+1)-dimensions.\footnote{Of course, there are universal effects
that can be studied around 1st order transitions as well, such as
finite-volume scaling.} 

Since the deconfinement phase transition of pure $SU(2)$ gauge theory
is interesting for theoretical purposes alone anyway, one might just
as well test the universality conjecture in other dimensions. This gives
much more freedom, since $SU(N)$ deconfinement phase transitions in
(2+1)-dimensions appear to be continuous at least for $N = 2$
\cite{Jesper1,Teper} and $N=3$ \cite{Jesper2} (and
probably for a few higher values of $N$). In (2+1) dimensions one
has the additional bonus that conformal symmetry in the corresponding
2-dimensional $Z(N)$ spin theory provides many more detailed
universality predictions \cite{Jesper3}. Again using the Polyakov
line in the fundamental representation of the gauge group, all results
have been consistent with the universality hypothesis also in (2+1)
dimensions.

There have, however, been some indications that the full picture may
be less simple. Suppose we ask for the critical dynamics of Polyakov
lines in other representations of the gauge group. If we associate
the Polyakov line in the fundamental representation with the $Z(N)$
spin variable, what are in spin-model language the analogues of the
Polyakov lines in higher representations? The behavior under $Z(N)$
transformations is supposed to give the answer. Thus, for representations
transforming trivially under $Z(N)$ there should be no critical dynamics
at all, since the associated Polyakov lines are not order parameters
of the phase transition. These lines should, for continuous transitions,
go smoothly from non-vanishing values in the confined to perhaps 
somewhat higher values in the deconfined phase. Representations that do
transform under $Z(N)$, on the other hand, are equally good order
parameters. From that point of view there should be no compelling
reason for singling out the fundamental representation as the analogue
of the spin operator. The only way to avoid a contradiction would seem
to be that all these representations should display the same critical
behavior near $T_c$ as the fundamental representations. 

To what extent are these general arguments supported by Monte Carlo
results? A clear change in behavior of the adjoint Polyakov line
at the deconfinement ``phase transition''(there is of course no genuine
transition in a finite volume) as measured on small lattices \cite{me1}
was already one indication that there could be difficulties with numerical
investigations of this problem. With the same level of statistics
(and the same lattice sizes) that were used routinely to confirm the
universality arguments based on the fundamental Polyakov line, 
a surprisingly different behaviour was found for the higher representations
of $SU(2)$ lattice gauge theory in (3+1) dimensions\cite{me2,Redlich}.
(For the continuous deconfinement transitions of $SU(2)$ and $SU(3)$
lattice gauge theories in (2+1) dimensions, see ref. 
\cite{Jesper1,Jesper2}). These numerical simulations indicated that sources
of higher representations that were sensitive to $Z(N)$ would correspond
to {\em different} magnetization exponents, one exponent for each
irreducible representation. But these results could all be criticized 
\cite{me2,Jesper1} on the grounds that they also seemed
to indicate critical behavior for Polyakov lines that simply are
not order parameters for the transition, those of transforming
trivially under $Z(N)$.
Indeed, in a Monte Carlo study of (3+1)-dimensional $SU(2)$ lattice gauge by 
Kiskis \cite{Kiskis1} the expected behavior (adjoint Polyakov line 
non-vanishing across the transition, the isospin 3/2 Polyakov line
behaving like the fundamental) was eventually extracted very close
to the finite-volume ``critical point'' $T_c$. 
In hindsight, the difficulty with higher representation sources was
perhaps to be expected. It had been observed earlier
that also  representations transforming trivially under the
center group could feel a linearly rising potential (with a slope
different from the string tension of the fundamental representation)
at intermediate distances \cite{Ambjorn}. At a certain range of distance
scales all representations appear to carry with them their own dynamics.
In the limit of an infinite number of colours, factorization is sufficient
to show that {\em all} irreducible representations are confined by
a linearly rising potential if the fundamental representation is, with 
string tensions that depend on the
representations \cite{Greensite}. Essentially, the intermediate distance
region in which a non-zero string tension exists for all representations
grows with $N$, the number of colours, reaching infinity as $N \to \infty$.
So for the relatively small latices sizes available, one is almost
always probing the intermediate distance regime, where all representations
feel a linearly rising potential -- a ``confining string''.

Conventional wisdom has it that the Gaussian fixed point with mean-field
exponents is the relevant one for all dimensionalities $d$ between a
certain specified ``upper critical dimension'' $d_u$ and $d = \infty$.
We shall not try to contradict the validity of this statement, but we shall
show that although it is an exceptional case,
it does {\em not} exclude the possibility of finding new
and unexpected critical behavior around this Gaussian fixed point.

It is important to realize that all of these issues can be addressed
even in the strong-coupling region of the lattice theory. In fact, in this
regime the universality arguments are even strengthened
(since the effective Polyakov-line interactions can be shown explicitly
to be short-ranged \cite{Polonyi,Green}), and the question of the critical
behavior of higher-representation
sources near the phase transition point is as meaningful in the 
strong-coupling regime as near the continuum limit. The advantage of
going to the strong-coupling regime is of
course that the question here can be studied in a much simplified setting
which still captures all the essentials. The leading-order effective 
Polyakov-line action reads \cite{Green}:
\beq
S_{eff}[W] = \frac{1}{2}J \sum_{x,j} \left\{Tr_F W(x) Tr_F W^\dagger(x+j) 
+ Tr_F W^\dagger(x) Tr_F W(x+j)\right\}~.
\eeq
Here the sum on $j$ runs over nearest neighbours.
The effective coupling $J$ is related to the gauge coupling $g$ and
$N_\tau$, the number of time-like links in the compactified temporal
direction. To lowest order, for $SU(2)$, it is 
\beq
J(g,N_\tau) = \left(\frac{I_2(4/g^2)}{I_1(4/g^2)}\right)^{N_{\tau}} ~,
\eeq
with $I_n$ indicating the $n$th order modified Bessel function.
For $SU(2)$ we will use a notation
in which $Tr_n W$ means the trace taken in the representation of isospin
$n/2$. $Tr_1 W$ is thus $Tr_F W$, and 
$Tr_2 W$ is the trace in the adjoint representation, etc. 
Higher orders in the 
expansion (1) (and corrections to the effective coupling (2)) can be
computed in a systematic expansion \cite{Gross2}, but we will not need
these corrections for the present purpose. The effective action (1)
becomes asymptotically exact in the strong coupling limit. 

It is useful 
to write
the effective Polyakov-line action (1) for $SU(2)$ in terms of a new
variable $\Phi(x) \equiv \frac{1}{2} Tr_1 W(x)$. The partition function
then takes the following form:
\beq
{\cal Z} = \int_{-1}^{1}[d\Phi] \exp
\left[ 4J\sum_{x,j} \Phi(x)\Phi(x+j) + \sum_x \tilde{V}[\Phi^2] \right] ~,
\eeq
with a local potential $
\tilde{V}[\Phi^2] = \frac{1}{2}\ln\left[1 - \Phi(x)^2\right]$.

There are two simple limiting cases in which the effective Polyakov-line
action (1) can be solved exactly. One is the large-$N$ limit \cite{me3},
where the deconfinement phase transition turns out to be of first order
(in agreement with large-$N$ reduction arguments based directly on
the full gauge theory \cite{Gocksch}), and where universality arguments 
hence cannot be addressed.\footnote{Still, the large-$N$ solution does
display a number of interesting features such as the simultaneous
deconfinement of all higher representations at the transition temperature
$T_c$, independently of whether these transformations transform trivially
under the center symmetry (in this case $U(1)$) or not. All representations
of the Polyakov line are hence equally good order parameters in this special
case, and all display a discontinuous jump at the phase transition.} The 
other exactly solvable case is the mean-field limit in
which $d \to \infty$, with $d$ being the number of spatial dimensions
\cite{me2}. In this limit one finds a genuine second-order critical point 
for the gauge group $SU(2)$ at a critical coupling $J_c \to 0$ as 
$d \to \infty$. For $J > J_c$ all higher-representation 
expectation values $\langle Tr_n W \rangle$ are non-zero \cite{me2}:
\beq
\langle Tr_n W \rangle ~=~  (n + 1) \frac{I_{n+1}(2a)}{I_1(2a)} ~,
\eeq
with $a = 2dJ\langle Tr_1 W\rangle$ being the self-consistent 
mean-field solution for 
the fundamental representation. Surprisingly, they {\em all} display
non-trivial critical behavior close to $J_c$:
\beq
\langle Tr_n W \rangle ~\sim~ (J - J_c)^{\beta_{n}} ~,
\eeq
where $\beta_n = n/2$. For the fundamental representation this just
corresponds to the mean-field Ising magnetization exponent $\beta_1
= \beta = 1/2$, in complete agreement with the universality arguments.

For the higher representations this new critical behavior is highly
unexpected. In this unphysical, but exactly solvable,
limit all standard screening arguments appear to break down, and we are
seeing new behaviour which is not predicted by universality.\footnote{One
also finds separate exponents $\delta_n\!=\!3/n$ for the behaviour of
$\langle Tr_n W\rangle\!\sim\!h^{1/\delta_{n}}$ at $J\!=\!J_c$ in a small
magnetic field coupled to $Tr_1 W$.}

It is
normally assumed that the relevant $Z(2)$ spin system universality class
to which the $SU(2)$ finite-$T$ phase transition should belong (if
continuous) would
display ``classical'' mean field exponents all the way down from $d\!=\! 
\infty$ to the upper critical dimension $d_u$ (in this case with 
$d_u = 4$, the critical behaviour being modified by logarithmic corrections
just at $d= d_u$). At a first glance this might seem to indicate that
the non-trivial behaviour (5) for {\em all} representations should
remain valid for all $d > 4$ ($d\!=\!4$ just being the limiting case), with 
non-trivial critical scaling even
for representations of integer isospin, and with new critical exponents
for all isospin half-integer representations as well. The first conclusion
simply cannot be correct (and we will show explicitly below why the argument
is invalid), because at strong coupling one can compute, for example, the
adjoint Polyakov line and see that to first non-trivial order in
$1/g^2$ it is non-zero. What about
the {\em odd}-$n$ representations? Could it be that they display new
non-trivial critical behaviour of the kind (5) even at $d\!=\!4$? 
Getting accurate Monte Carlo results close to $T_c$ in the full 
$SU(2)$ gauge theory is extremely difficult and time-consuming.
To circumvent this problem, a Monte Carlo simulation has instead been
performed on 
the effective Polyakov-line action for $SU(2)$ in $d\!=\!4$ spatial
dimensions. From a conceptual point of view this is no restriction at
all, since -- as we have emphasized before -- all these questions are
as important in the context the strong-coupling effective 
action (1) as in the full gauge theory.

The critical coupling of the model was determined using the fourth order
cumulant \cite{Binder} 
\beq
U_1 = 1 - \langle m^4 \rangle\!/(3\langle m^2 \rangle^2) ~,
\eeq 
where $m$ is the 
single lattice average over $\Phi$, $i.e.$ the Polyakov line in the 
fundamental representation.
In these runs the lattice sizes were $L\!=\!4, 6, 8, 12$ and 16.
The resulting curves are plotted in Fig. 1. The crossings of the cumulant
provide estimates of the critical coupling $J_c$. The error vanishes
like $L^{-1/\nu}$. The crossings of the Binder cumulant for 
$L=12$ and $L=16$ lattices were taken as the 
best estimate of the critical coupling, giving
$~4 J_c = 0.5507(2)$.

Next, consider the fourth-order cumulant $U_n$ for higher representations. 
Fig. 2 shows the results for $n\!=\!2$.  
With increasing lattice sizes
the cumulant converges toward $2/3$ for couplings both below and above 
the critical point. The value $U\!=\!2/3 $ signals a finite expectation value 
of the observable: the Polyakov line in the adjoint 
representation is not an order parameter. The 
fourth-order cumulant for the $n\!=\!3$ representation takes values close
to $2/3$ in the broken phase and values close to zero in the high temperature
phase; it behaves as an order parameter. But the curves do not 
display crossings 
close to the critical coupling predicted from the cumulant for the fundamental
representation. The curve for $L\!=\!16$ comes close to that of the fundamental
representation, so we might expect that for still larger lattices the cumulant
for the $n\!=\!3$ representation converges towards the fundamental ones, 
and that the 
crossings can then be observed at the critical coupling $J_c$. 
The data for the higher representation were so affected by errors that 
no reliable results for the cumulants could be extracted.

The model was also simulated for various $J > J_c$ on lattices of 
sizes up to $L\!=\!16 $, the aim being an approximate determination of 
the critical exponents $\beta_n$ 
directly from the Monte Carlo measurements of the different representations.  
A $J$-dependent  
``effective''  exponent $\beta^{eff}_n$ can be defined by 
\begin{equation} 
\beta^{eff}_n = (J-J_c) \frac{d\langle Tr_n W \rangle/dJ}{\langle Tr_n 
W \rangle}.
\end{equation} 
The derivative of $\langle Tr_n W\rangle$ with respect to $J$ can then be
computed from the relation
\begin{equation} 
\frac{d}{dJ}\langle Tr_n W \rangle = \langle (Tr_n W)\cdot \tilde{S}\rangle
- \langle Tr_n W \rangle \langle \tilde{S} \rangle~.
\end{equation} 

The final results for this are presented in fig. 4 (for the odd 
representations) and 
fig. 5 (for the even ones). Only values which were consistent on the 
two largest lattice sizes were considered. It turned out that for the 
coupling close
to $J_c$ even the $L=16$ lattice was not sufficient to give a stable result
for the $n=5$ representation. The curves plotted in these two
figures are improved mean-field predictions. They will be explained in
detail below. 

Figs. 4 and 5 demonstrate fairly convincingly that the odd representations
converge toward an effective $\beta_n^{eff} = 1/2$ independent of $n$, while
the even representations converge toward $\beta^{eff}_n = 0$ (as expected
if these representations remain finite at $J_c$). But the plots also 
reveal an interesting phenomenon for larger values of $(J - J_c)/J_c$:
the effective $J$-dependent exponents $\beta^{eff}_n$ quickly
reach a regime of couplings where they are essentially equally spaced, 
growing linearly with $n$. Although they never actually reach the 
mean-field prediction (5), they get quite close, and they certainly
obey the rule $\beta^{eff}_n \sim n\cdot\beta^{eff}_1$ to surprisingly
high accuracy. This is just as for the original observations in the full 
$(3\!+\!1)$-dimensional $SU(2)$ gauge theory \cite{me2,Redlich}. It appears
that this approximate linear relation between the $\beta_n$'s, when 
measured not too close to the critical point, can be viewed as the 
``remnant'' of the $d\!=\!\infty$ solution. It is then only {\em very} close
to the critical point the behaviour changes, and the single critical
exponent $\beta$ emerges for the odd representations, while the even
representations run smoothly across the transition point. We can estimate
this narrow window in the original gauge coupling $4/g^2$ by using the
relation (2). In the case of $N_{\tau} = 2$ the transition occurs at
$4/g_c^2 = 1.6424(4)$. In order to obtain $\beta_{eff} < 0.625$ (i.e.
25$\%$ above the correct value $\beta=0.5$) 
for the $n\!=\!3$ representation we would have to take $4/g^2 < 1.66$.

While these results may have clarified the situation in the $d\!=\!4$ theory,
we are still left with the surprising $d\!=\!\infty$ results where mean field
theory is believed to be exact. How can they be explained? Consider the
representation of the effective Polyakov-line action given in eq. (3).
This is a $Z(2)$-invariant effective scalar field theory in $d$ dimensions,
as expected on general grounds. But it is a {\em very particular} effective
scalar theory, one that embodies the underlying $SU(2)$ structure (in the
restrictions on the integration interval of $\Phi(x)$, and in the very
special form of the local potential $\tilde{V}[\Phi^2]$, which reflects 
the Haar measure for $SU(2)$).

Since we at this point wish to focus on the $d = \infty$ results,
we can restrict ourselves to ``classical'' mean-field considerations. It
is instructive \cite{Jesper1} to generalize the partition function above
to an arbitrary local potential $V[\Phi^2]$ and relax the limitation
on the integrations over $\Phi(x)$ to be in the interval $[-1,1]$. The
$d = \infty$ solution is then found by considering the single-site
partition function
\beq
{\cal Z}_{SS}~=~\int_{-\infty}^{\infty}[d\Phi]\exp\left[v\Phi 
+ V[\Phi^2]\right] ~,
\eeq
where $v = 4dJ\langle\Phi\rangle$ will be determined by the 
self-consistency solution.
Clearly, for $n$ being any non-negative integer, $\langle\Phi^{2n+1}
\rangle = 0$ unless 
the global $Z(2)$ symmetry is spontaneously broken. Call the critical 
coupling at which this occurs $J_c$. If the phase transition is continuous, 
$\langle\Phi\rangle$ will be small just above $J_c$, and it is meaningful 
to expand in 
$v$ (no matter how large $d$ is taken, once fixed). The result is, for the 
expectation values of the first two non-trivial mean-field moments of 
$\Phi$ \cite{Jesper1}:
\begin{eqnarray}
\langle\Phi^2\rangle &=& \langle\Phi^2\rangle_0 + \frac{1}{2}\left[\langle
\Phi^4\rangle_0 - \left(\langle\Phi^2\rangle_0
\right)^2\right] v^2 + \ldots \cr
\langle\Phi^3\rangle &=& \langle\Phi^4\rangle_0 v + \frac{1}{2}\left[
\frac{1}{3}\langle\Phi^6\rangle_0
- \langle\Phi^2\rangle_0\langle\Phi^4\rangle_0\right] v^3 + \ldots ~,
\end{eqnarray}
where the subscript ``0'' indicates the (constant) expectation value in 
the unbroken phase $J < J_c$.
Higher moments can be worked out analogously, by expanding both
the partition function ${\cal Z}_{SS}$ and the unweighted averages 
in powers of $v$. Using the recursion relation $\chi_{n+1} = \chi_n\chi_1
- \chi_{n-1}$ for $SU(2)$ characters, we find the general $d = \infty$
predictions \cite{Jesper1}
\begin{eqnarray}
\langle Tr_2 W\rangle &=& \left[4\langle\Phi^2\rangle_0-1\right] + 
2\left[\langle\Phi^4\rangle_0
- \left(\langle\Phi^2\rangle_0\right)^2\right] v^2 + \ldots \cr
&=& A_2 + B_2 v^2 + \ldots \cr
\langle Tr_3 W\rangle &=& \left[8\langle\Phi^4\rangle_0 - 4\langle
\Phi^2\rangle_0\right] v +
4\left[\frac{1}{3}\langle\Phi^6\rangle_0 - \langle\Phi^2\rangle_0
\langle\Phi^4\rangle_0 + \frac{1}{2}
\langle\Phi^2\rangle_0\right] v^3 + \ldots \cr &=& A_3 v + B_3 v^3 + 
\ldots ~,
\end{eqnarray}
where $A_2, B_2, A_3$ and $B_3$ are (non-universal) constants.
This shows the behaviour expected from universality arguments. The adjoint
Polyakov line will remain non-vanishing across the phase transition at
$J_c$ (and is hence not an order parameter), and the isospin-3/2 
representation scales near $J_c$ as $v$, $i.e.$, as the fundamental
representation. But if we take the particular potential $\tilde{V}[
\Phi^2]$ of eq. (3), and restrict the integration over $\Phi$ to the
interval $[-1,1]$, then devious cancellations occur. One finds 
$\langle\Phi^2\rangle_0 = 1/4$ and $\langle\Phi^4\rangle_0 = 1/8$, leading to
\begin{eqnarray}
\langle Tr_2 W\rangle &=& \frac{1}{8} v^2 + \ldots = 
2d^2J^2\langle\Phi\rangle^2 + \ldots \cr
\langle Tr_3 W\rangle &=& \left[\frac{4}{3}\langle\Phi^6\rangle_0 + 
\frac{3}{8}\right] v^3
+ \ldots ~.
\end{eqnarray}
It is thus suddenly the {\em non-leading} terms in the general expansion
of the Polyakov lines that become important, due to the amplitudes
of the leading terms vanishing in this limit.
The cancellations required for this phenomenon are actually simple 
consequences of the orthogonality
relations for $SU(2)$ characters, as follows if one performs the
mean field calculation directly in $SU(2)$ language \cite{me2}. They
occur similarly for all higher representations, leading, of course,
eventually to the general $d\!=\!\infty$ solution (5).

We are now in a better position to understand the $d = \infty$ results.
As shown above, the appearance of
new exponents for each of the odd-$n$ representations in the limit
$d\!=\!\infty$ is due to very delicate cancellations that make the
{\em amplitudes} of the leading terms in the expansion close to the critical
point vanish. Although the same mechanism is responsible for the fact
that also even-$n$ representations display non-trivial
critical behaviour in the $d\!=\!\infty$ theory, that phenomenon is of course
far more difficult to understand from the point of view of physics.   
The even-$n$ Polyakov-line representations simply ought not to be order
parameters for the deconfinement transition, even in the $d\!=\!\infty$ limit,
since such sources should be screened both above and below the
critical point. The resolution of this apparent paradox lies in the fact
that the critical coupling $J_c$ actually {\em vanishes} (like $1/d$) when
$d\!\to\!\infty$, as follows directly from the mean-field solution 
(4). This behaviour is not an artifact of the mean-field solution;
it can be checked to hold as well in the exact solution of the  
$N = \infty$ theory \cite{me3}. In terms of the gauge coupling $g$ this
entails, for fixed $N_\tau$, $g\!\to\!\infty$. Although 
this makes the 
strong-coupling effective Lagrangian analysis more and more accurate,
it also pushes the confinement/deconfinement phase transition right to
the extreme limit $g\!=\!\infty$ where {\em all} sources are ``confined''
($\langle Tr_n W\rangle = 0$ for all $n$ at $g\!=\!\infty$ in the full 
gauge theory
simply as a consequence of
the orthogonality property of the group characters). It is for this simple
reason that the mean-field solution, correctly, predicts critical behaviour
for all representations of $SU(2)$.

\section{Improved Mean Field Theory}

The limit $d\!=\!\infty$ of finite-temperature gauge theories is thus in
many respects highly singular. This, together with the Monte Carlo data
presented above for the $d\!=\!4$ $SU(2)$ theory, indicates that the usual
assumption of $d\!=\!\infty$ exponents being valid down to the upper critical
dimension $d_u$ simply fails in this case. Can we understand the 
singular nature
of the $d\!=\!\infty$ limit in an analytical way? As explained above,
there are actually no reasons to doubt that mean field theory predicts
the $d\!=\!\infty$ behaviour correctly. The only resolution would then be
that {\em any} finite dimensionality $d$ should correspond to radically 
different behaviour close to the critical point, $i.e.$, that 
$1/d$-corrections discontinuously should alter the critical indices. To
see whether this is the case, we have considered a slightly improved
mean-field solution of the same effective Polyakov-line action (1). (This
improvement appears to be equivalent to what is known as the 
Bethe-approximation, see, $e.g.$, ref. \cite{bethe}).
 
The Bethe-approximation can be seen as the lowest order improvement in a 
whole class of mean-field improvements. 
Let us consider a lattice field theory defind  by the action
\begin{equation}
S=-\beta \sum_{<xy>} \phi_x \phi_y  + \sum_x V(\phi_x) \;
\end{equation}
where we assume for simplicity of the argument that $ \phi$ is a real variable. 
 In standard mean-field approximation one replaces the neighbours of a 
 spin by an external field. 
\begin{equation}
 Z_{M}= \int \mbox{d} \phi  \exp(\beta 2d H \phi -V(\phi))
\end{equation}
The value of the external field is fixed by the self-consistency condition
$ H =  \langle \phi \rangle (H) $. 

The idea to improve the mean-field approximation 
 is to consider
a finite lattice rather than a single site. Again the missing neighbours 
of the spins at the boundary are replaced by an external field, which just
acts on those spins at the boundary.  
\begin{equation}
 Z_{IM}= \int [\mbox{d} \phi]  \exp(+ \beta \sum_{<xy>} \phi_x \phi_y
+ \sum_x \beta n_x H \phi_x  -\sum_x V(\phi_x))
\end{equation}
where  $n_x$ is the number of missing neighbours  at the site $x$. 
Having a larger lattice there is some ambiguity in the choice of the 
self-consistency condition. 
Assuming that the best approximation of the true
physics is obtained in center of the lattice we require 
$  \langle \phi_c \rangle (H) = \langle \phi_{nc} \rangle (H) $.
where $c$ denotes a site at the center of the lattice and $nc$ next to the 
center. 
It turns out that the critical exponents obtain from this type of improved
mean-field are in general identical to those of standard mean-field. However
non-universal quantities like off-critical expectation values and the 
critical temperature converge systematically towards the  true values as 
the size of the lattice increases. This behaviour can be nicely illustrated 
at the example of the two dimensional Ising model.
One obtains from square lattices of size $L=2,3,4$ and 5 the inverse critical
temperatures $\beta_c =$ $ 0.285745...$, $0.380902...$ , $0.388503...$  and
$0.400785...$ respectively.  These numbers have to be compared with the 
exact solution $\beta_c = 0.4406868...$ and the standard mean-field result
$\beta_c=0.25$.
For a discussion of even more general improved mean-field  methods see ref. 
\cite{Suzuki94} and references therein. 

Relying on standard numerical integration for a continuous field $\phi$ only  
the smallest lattices are managable. In our study we considered a lattice
consisting out of a central site (C) and its $2d$ neighbours (O).
 Due to factorization
this problem is not harder to solve than a two-site problem. 

In order to get a rough idea about the improvement that can be obtained  we 
computed the "star"-approximation for the  3D Ising model on a simple cubic
lattice. From standard mean-field one gets $\beta_c=1/6=0.16...$. The 
improvement gives $\beta_c=0.202732...$ , which has to be compared 
with the MC-estimate $\beta_c= 0.221652(3)$ \cite{baillie}. 


The partition function of this system is given by
\begin{equation}
Z = \int_{-1}^{+1} d\Phi_C \sqrt{1-\Phi_C^2}
     \prod_O \int_{-1}^{+1} d\Phi_O \sqrt{1-\Phi_O^2}
        exp(4 J (\Phi_C + W)  \Phi_O)
\end{equation}
in the case of the effective  Ployakov-line action. 
The integration over the $\phi_O$ fields leads to  
\begin{equation} 
Z \propto \int_{-1}^{+1} d\Phi_C \sqrt{1-\Phi_C^2}
 \left[\frac{I_1(4 J (\Phi_C + W))}{4 J (\Phi_C + W)}\right]^{2d}
\end{equation} 
The remaining one-dimensional integration we have performed numerically.
In order to fix the external field $W$ we require, as above,
 that the magnetization
of the fundamental representation is equal for the central site (C) and its 
neighbours (O). 
Expectation values are evaluated on the central site. 
We have solved the self-consistency equations
for $d=4, 8, 16$ and 32.  The value found for the critical coupling $J_c = 
0.5352...$ in 4-$d$ deviates from the Monte Carlo 
value by only $2.8\%$, while standard mean field theory is off by $9.2\%$. 
But a more striking consequence of the improvement is seen in the behaviour
of the {\em even}-n representations, which are now non-vanishing for
all values of the coupling $J \neq 0$. The numerical values at or below
$J_c$, however,
decrease rapidly with $d$. For the adjoint representation it is reduced
by a factor of approximately 2 at $J_c$ when one doubles the dimension,
while for the $n\!=\!4$ representation it drops by almost a factor of 4. In
this fashion the present solution matches the usual mean-field results in
the limit $d\!=\!\infty$.

With the improved mean field theory we can finally make a much more accurate
comparison with our $d\!=\!4$ Monte Carlo results. In figs. 4 and 5 we have
thus plotted (as smooth curves) the corresponding predictions for the
$J$-dependent effective exponents $\beta^{eff}_n$ defined as in eq. (8).
Qualitatively the behaviour of the Monte Carlo data is quite well reproduced.

Clearly, as $d\!\to\!\infty$ the window in which the conventional
results are reproduced shrinks to zero. In fact, one can easily estimate
from the improved mean-field solution (16) that this window decreases in size
as $1/d$, eventually disappearing at $d\!=\!\infty$. In the more
conventional language, this is the point at which the amplitudes of
the leading terms in the expansion for the Polyakov lines vanish. 

\section{New Results for SU(3)}
The effective Polyakov-line action (1) for the finite temperature
$SU(3)$ gauge theory can be written in a similar form as (3). Any 
$U \in SU(3)$ is unitary equivalent to the diagonal matrix
 $diag (e^{i\phi_1},e^{i\phi_2},e^{i\phi_3})$ with 
$\phi_1+\phi_2+\phi_3=0 \;\; mod \;\; 2\pi$. The Haar measure takes the 
form 
\begin{equation}
\int dU ... = \frac{1}{6} \int_{0}^{2\pi} \frac{d\phi_1}{2\pi} \int_{0}^{2\pi}
\frac{d\phi_2}{2\pi}
 \prod_{1 \le k \le l \le 3} |e^{i\phi_k}-e^{i\phi_l}|^2 ...
\end{equation}
and the trace of the $n^{th}$ power of a SU(3) matix is given by
\begin{equation}
 s_n = Tr U^n = \sum_{j=1}^3 \exp(i n \phi_j) \;.
\end{equation}
The characters of the irreducible representations can be written as 
finite polynoms in the $s_n$ and their complex conjugate. 
In the following we shall consider
\begin{eqnarray}
\chi_1(U) &=& 1 \\
\chi_3(U) &=& s_1 \\
\chi_6(U) &=& \frac12 (s_1^2 + s_2) \\
\chi_8(U) &=& |s_1|^2 -1 \\
\chi_{10}(U) &=& \frac16 (s_1^2 + 3 s_1 s_2 + 2 s_3) 
\end{eqnarray}
where the index of $\chi$ gives the dimension of the representation. 
Since the $SU(3)$ finite temperature gauge theory is expected to undergo 
a first order phase transition, we have to compute the free energy to verify
this assumtion and to obtain the critical temperature. 
In order to extract a result from the 
star-lattice, we try to compensate for boundary effects in the optimal way.
In addition to the star lattice we consider a lattice consisting of just two 
sites. The star lattice has $2d$ internal links and $2d (2d-1) $ external
links. The two site lattice has 1 internal link and $2 (2d-1)$ external links.
The best approximation we expect from a linear combination where the 
external links are cancelled.  Therefore we chose
\begin{equation}
 f_{IM}(H) = - \ln Z_{star}(H) + d \ln Z_{2}(H)
\end{equation}
as approximation for the free energy. 

The value of the external field $H$ is then fixed by minimizing the free
energy with respect to $H$. The occurence of two minima means
that the phase transition is of first order.  
We find for d=3 $J_c \approx 0.1343$ and 
$J_c \approx 0.1369$ for mean-field and improved mean-field respectively. 
These results have to be compared with the Monte Carlo estimate $J_c=0.13722$
obtained  by K.Rummukainen  \cite{kari}.

In fig. 6 we give the expectation value of the Polyakov line for various
representations. We also give the solution for the second miniumum of the 
free energy. These results are relevant for a supercooling of the system. 
In contrast to the lowest-order mean field solution \cite{me2}, the 
expectation value of the Polyakov line in the adjoint representation  
does not vanish in the low-temperature phase.

\section{Abelian Projections}

\vspace{0.5cm}

\noindent  The abelian projection theory of quark confinement, put forward by
't Hooft in ref. \cite{tHooft}, is a proposed identification of the
physical degrees of freedom that are 
the most relevant to the infrared dynamics.  The idea
is based on a gauge choice, which reduces the underlying $SU(N)$ gauge
symmetry to an abelian $U(1)^{N-1}$ symmetry generated by the Cartan
subalgebra of the gauge group.  The non-abelian theory, with such a gauge
choice, resembles a generalization of compact QED, in which (abelian) charged
quarks and gluons interact by an exchange of "photons," 
which are the (abelian) neutral gluons corresponding to generators of the
Cartan subalgebra.  
The abelian gauge field also contains monopoles, which may
be identified from certain degeneracies in the gauge-fixing condition.
At this stage, the identification of abelian neutral gauge fields as the
"photon" fields of an abelian gauge theory, with remaining gauge fields 
labeled as abelian charged vector bosons, is simply kinematics.

   The hypothesis of the abelian projection theory is that
it is the photon/monopole fields, i.e. the abelian gauge fields, 
that are the crucial degrees of freedom 
with respect to quark confinement.  Confinement, in this picture, 
is due to condensation
of monopoles associated with the $U(1)^{N-1}$ gauge fields, as in compact
$QED_3$ \cite{Poly}.  Now, if the crucial
degrees of freedom are those of an abelian gauge theory, it is 
reasonable to suppose that the vacuum fluctuations of these abelian gauge 
fields would dominate the vacuum fluctuations of the theory at large scales.  
Suppose, for example, that the gauge group is $SU(2)$, and the gauge 
choice is the maximal abelian gauge, which (on the lattice) maximizes 
the quantity
\beq
       Q = \sum_{x,\mu} \mbox{Tr}[U_\mu (x) \sigma^3 
              U^\dagger_\mu (x) \sigma^3]
\eeq
There is then a remaining $U(1)$ symmetry, and the corresponding
"abelian" gauge field is $A_\mu^3(x)$.  If it is the vacuum fluctuations in
$A_\mu^3$ which dominate at large scales, then it might be a reasonable
approximation, e.g. for purposes of calculating large Wilson loops or
perhaps also
Polyakov lines, to ignore the contribution from the other color 
components, i.e.
\bea
       \langle W_j(C)\rangle &=& {1\over 2j+1}\langle
\mbox{Tr}\exp[i\oint dx^\mu A^a_\mu T^j_a]\rangle
\nonumber \\
              &\sim& {1\over 2j+1}\langle
\mbox{Tr}\exp[i\oint dx^\mu A^3_\mu T^j_3]\rangle
\label{abdom}
\eea
where $T^j_a$ are the $SU(2)$ generators in the $j$-representation, and we
normalize $W_j$ to a maximum value of $1$.
This approximation was originally applied by Polyakov,
to calculate Wilson loops in the D=3 dimensional Georgi-Glashow 
model in the Higgs phase \cite{Poly}.  In the Georgi-Glashow model,
the $U(1)$ symmetry 
is singled out by a unitary gauge choice.  In the context of the abelian 
projection theory, the approximation is known as "abelian 
dominance." \cite{Suzuki1}  The validity of abelian dominance is rather
important to the abelian projection theory, because if this 
approximation turns out to be wrong, if in fact
it is {\it not} true that fluctuations in $A_\mu^3(x)$ dominate the
vacuum at large scales, then it is unclear in what respect
the monopole/photon degrees of freedom of the abelian projection theory
are the most crucial for quark confinement.

   On the lattice, abelian dominance is tested by calculating,
via lattice Monte Carlo,
Wilson loops obtained from "abelian-projected" link configurations 
in maximal abelian gauge.  In $SU(2)$
gauge theory this means setting the off-diagonal components of the
gauge-fixed $U_\mu(x)$ matrices to zero,
and then rescaling each link by a constant to restore unitarity.    
Abelian dominance seems to work quite 
well, in numerical simulations, for extracting
string tensions from Wilson loops in the fundamental representation
\cite{Suzuki1}.  Very recently, however, it has been shown 
in ref. \cite{anti} that, for $SU(2)$ gauge theory in $D=3$ dimensions, 
abelian dominance fails entirely when applied
to Wilson loops in higher group representations.  This failure is
connected with the "Casimir scaling" of interquark forces.
It is well known
that the force between heavy quarks
in any group representation, in a distance interval from onset of 
confinement to the onset of color screening, is proportional to
the quadratic Casimir of the representation.  This fact has been seen
both in $D=3$ and $D=4$ dimensions, for $SU(2)$ and $SU(3)$ gauge 
groups \cite{Ambjorn,Casimirs}.
Moreover, the interval between the onset of confinement and
the onset of screening is quite large in the scaling region; it is not
even clear that color screening has been seen yet in $D=3$ dimensions
(c.f. Poulis and Trottier in \cite{Casimirs}).  For $SU(2)$, the
numerical results in this interval, in accord with Casimir scaling,
are that the $j={3\over 2}$ tension is about 5 times the $j={1\over 2}$ 
tension, and the $j=1$ tension is about ${8\over 3}$ the $j={1\over 2}$
string tension \cite{Ambjorn,Casimirs}.  In contrast to this
Casimir scaling of interquark forces, it was
found in ref. \cite{anti} that string tensions extracted from loops built
from abelian projected configurations are roughly equal for the $j=1/2$
and $j=3/2$ representations, and are consistent with zero
for $j=1$.  For the $j>{1\over 2}$ representations, abelian dominance
is not even approximately true.

    There is ample motivation, then, to critically examine the
hypothesis of abeliance dominance in the case of 
Polyakov lines in various representations.
Again, the abelian-projected link variables in $SU(2)$ gauge theory
are constructed by truncating the full link variable
\beq
         U = a_0 I + i\sum_{k=1}^3 a_k \sigma^k
\eeq
to the diagonal component, followed by a rescaling, i.e.
\beq
         U \rightarrow U' 
           = {a_0 I + i a_3 \sigma^3 \over \sqrt{a_0^2 + a_3^2} }
\label{abpro}
\eeq
The abelian links $U'$ are then used in the computation of the Polyakov
lines.

   We work in maximal abelian gauge.  In this gauge there is nothing
special about the time direction; string tensions 
have been extracted from loops in all space-time
orientations.  For spatially asymmetric gauges, there is a danger that
abelian dominance (even for fundamental representations) may only work
for loops oriented in certain directions, as found by one of us (J.G.) and 
Iwasaki in ref. \cite{anti1}. 

   Fig. 7 shows the Polyakov lines in the fundamental representation,
computed using the full and abelian-projected configurations.
The lattice size was $12^3\times4$, and shown are raw data before
averaging, after 500 iterations. The errors are insignificant in this
context, since we are not interested in a very detailed comparison
of numbers close to $T_c$. We have indicated the full Polyakov lines
by $W$, and the abelian-projected Polyakov lines by $V$, both normalized
to a maximum value of 1. The results are similar
to those found in ref. \cite{Hioki}.  There is clearly
a steep rise in Polyakov line values at the deconfinement transition,
for both the full and abelian-projected configurations.  There is,
just as clearly, a {\it quantitative} disagreement in the full and
abelian-projected results.  The disagreement becomes much more
pronounced for Polyakov lines at $j=1$, shown in Fig. 8; in this case 
it is not even clear that we have qualitative agreement!  Note in
particular the fact that the value of the adjoint line formed from
abelian-projected configurations tends to a non-zero constant
(${1\over 3}$) at $\beta=0$.  Finally, the $j=3/2$ results are shown
in Fig. 8; again there is considerable quantitative disagreement
between the values of the full and abelian-projected lines.

   The origin of the asymptotic value of ${1\over 3}$, obtained as $\beta 
\rightarrow 0$ for the $j=1$
representation Polyakov line, is not hard to understand.  Computing adjoint
lines with abelian-projected configurations, we have
\bea
   \langle W_1(C)\rangle_{ab. proj.} 
         &=& {1\over 3}\langle\mbox{Tr}\exp[i\oint dx^\mu A^3_\mu T^1_3]
\rangle \nonumber \\
         &=& {1\over 3}\sum_{m=-1}^1 \langle(m|\exp[i\oint dx^\mu 
A^3_\mu T^1_3|m)\rangle \nonumber \\
         &=& {1\over 3}\sum_{m=-1}^1 \langle\exp[im\oint dx^\mu A^3_\mu]
\rangle \nonumber \\
         &=& {1\over 3} + {1\over 3}\langle\exp[i\oint dx^\mu A^3_\mu] 
                          + \mbox{c.c.}\rangle
\nonumber \\
         &\rightarrow& {1\over 3} ~~~~~\mbox{as}~~~ \beta \rightarrow 0
\eea
The adjoint representation of a Polyakov line built entirely out of 
abelian-projected fields has an abelian neutral 
component, corresponding to $m=0$ in the above sum, which
remains unaffected by the vacuum 
fluctuations (no matter how large) of the abelian gauge field.
The behavior of adjoint Polyakov lines  constructed
from the full, unprojected configurations is, of course, very different.
The value of adjoint Polyakov lines in the unprojected case
is related to the energy of a gluon bound to the adjoint source.
This energy becomes infinite, in lattice units, as $\beta \rightarrow 0$,
which is why the corresponding adjoint Polyakov line vanishes in that
limit.  The fact that the abelian-projected adjoint loop 
is finite (${1\over 3}$) in the same limit indicates
that the abelian dominance approximation is insensitive to the actual
dynamics of color screening.  It also indicates that for the full,
unprojected Polyakov lines, the vacuum fluctuations of the  "charged boson" 
fields, i.e. $A^1_\mu$ and $A^2_\mu$, must play an essential role.

   Evidently, the abelian projection approximation (eq. \rf{abpro}) in the
maximal abelian gauge is {\it not}
a very good approximation for Polyakov lines, with the 
quantitative disagreement between full and abelian projected values 
being especially severe for the higher ($j > {1\over 2}$) representations.
This disagreement is in line with the results for Creutz ratios found
in ref. \cite{anti} (certain other criticisms of the abelian projection theory
can be found in ref. \cite{anti1}) 
and in general suggests a failure of the abelian 
dominance approximation in maximal abelian gauge.  A failure of abelian 
dominance, in turn, would imply that an effective
abelian theory at the confinement scale, invoking only the monopoles
and photons identified by abelian projection, is inadequate to describe 
the actual non-perturbative behavior of non-abelian gauge theory. 

\vspace{0.5cm}

{\sc Acknowledgment:}~ We would like to thank K. Anagnostopoulos for
help in creating some of the figures. J.G.'s research is supported in part 
by the U.S. Dept. of Energy under Grant No. DE-FG03-92ER40711. 


\newpage

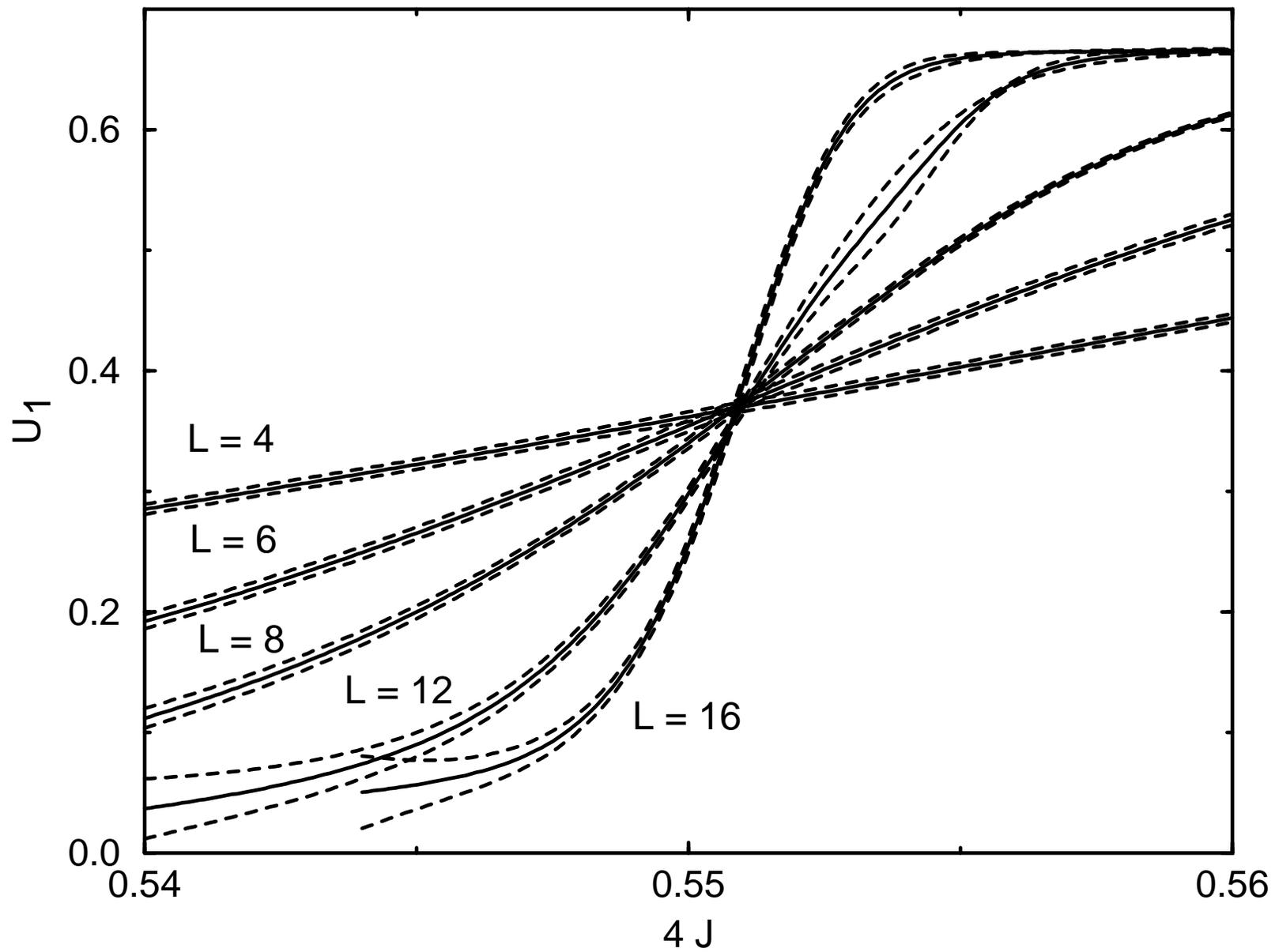
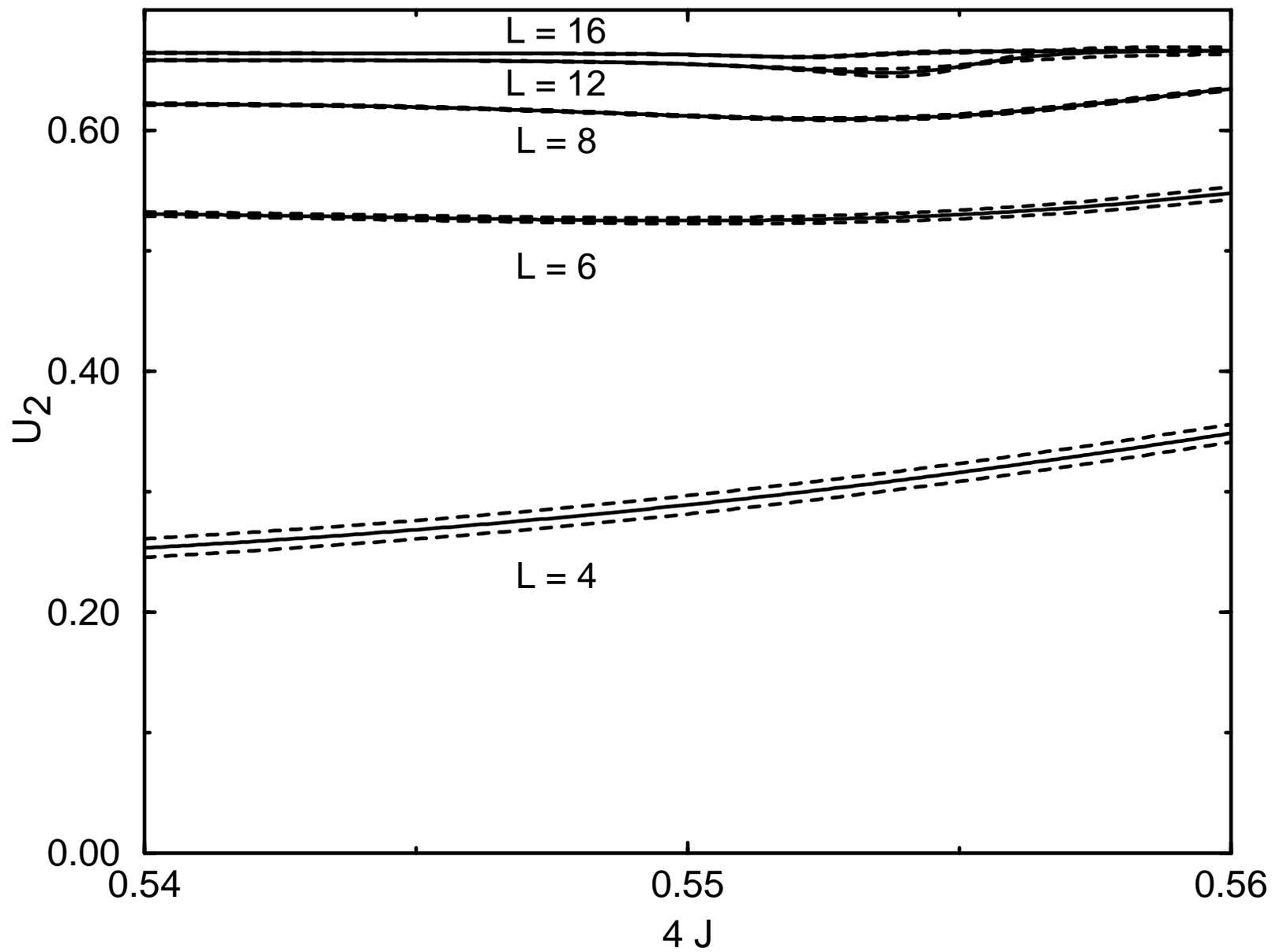
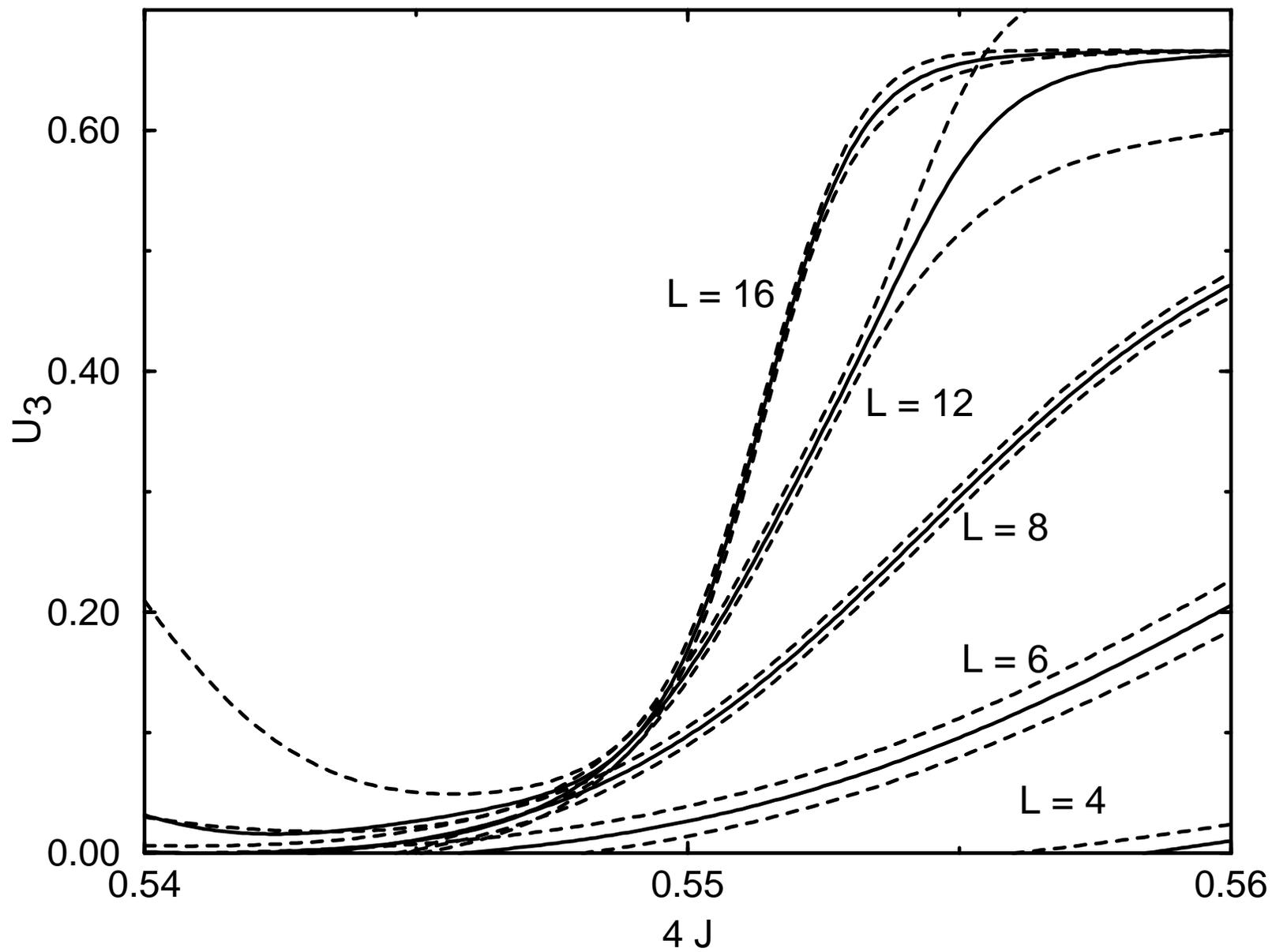
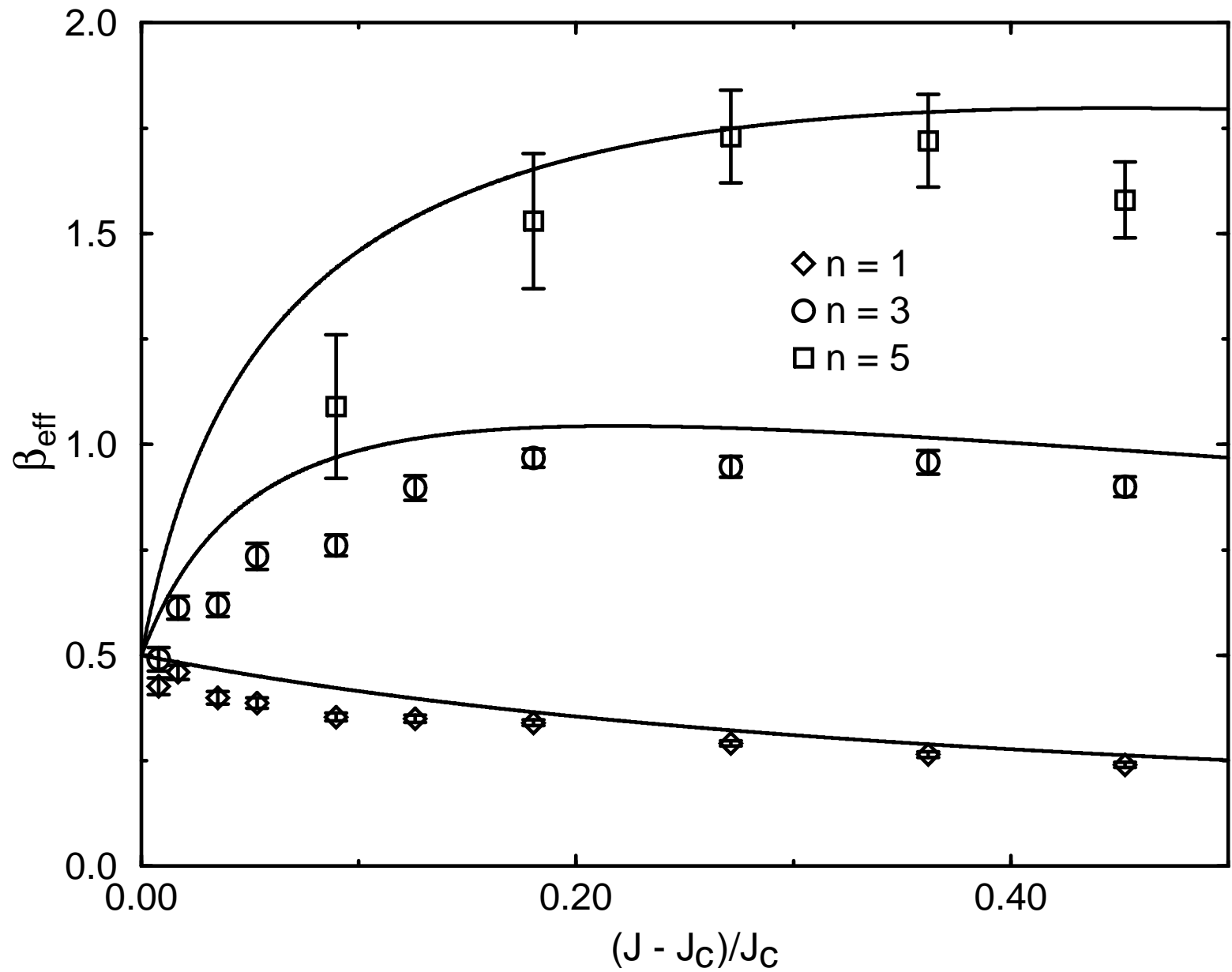
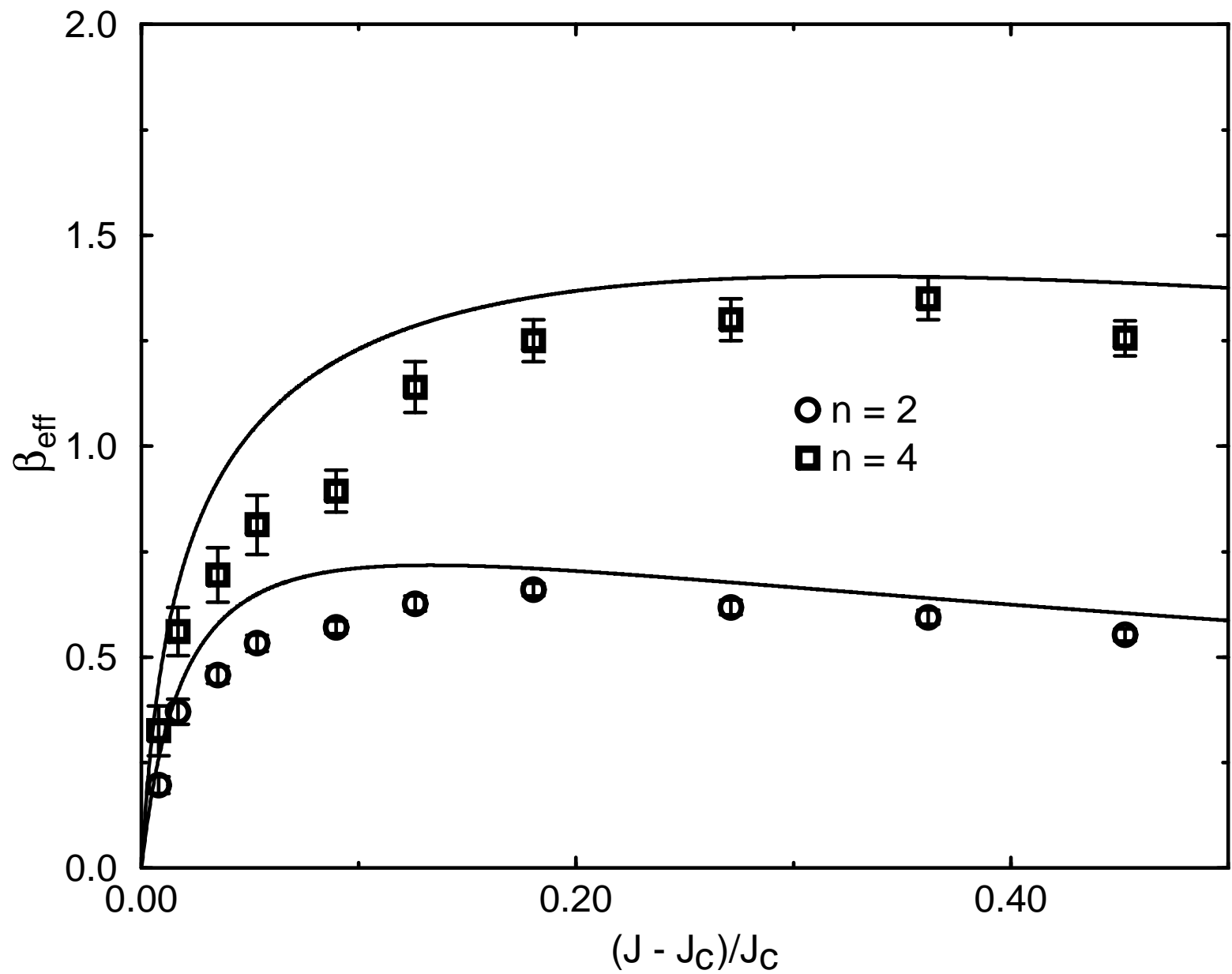
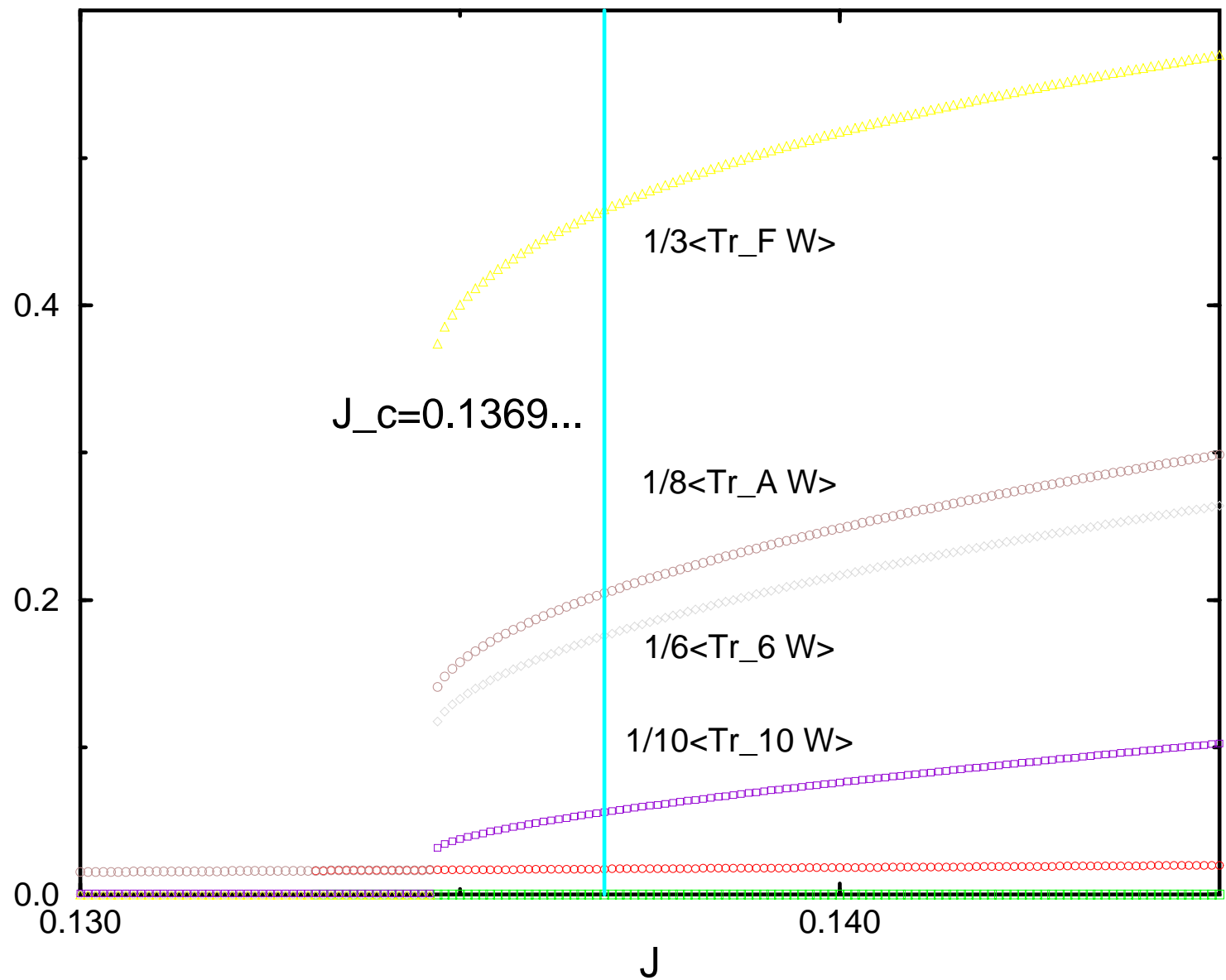
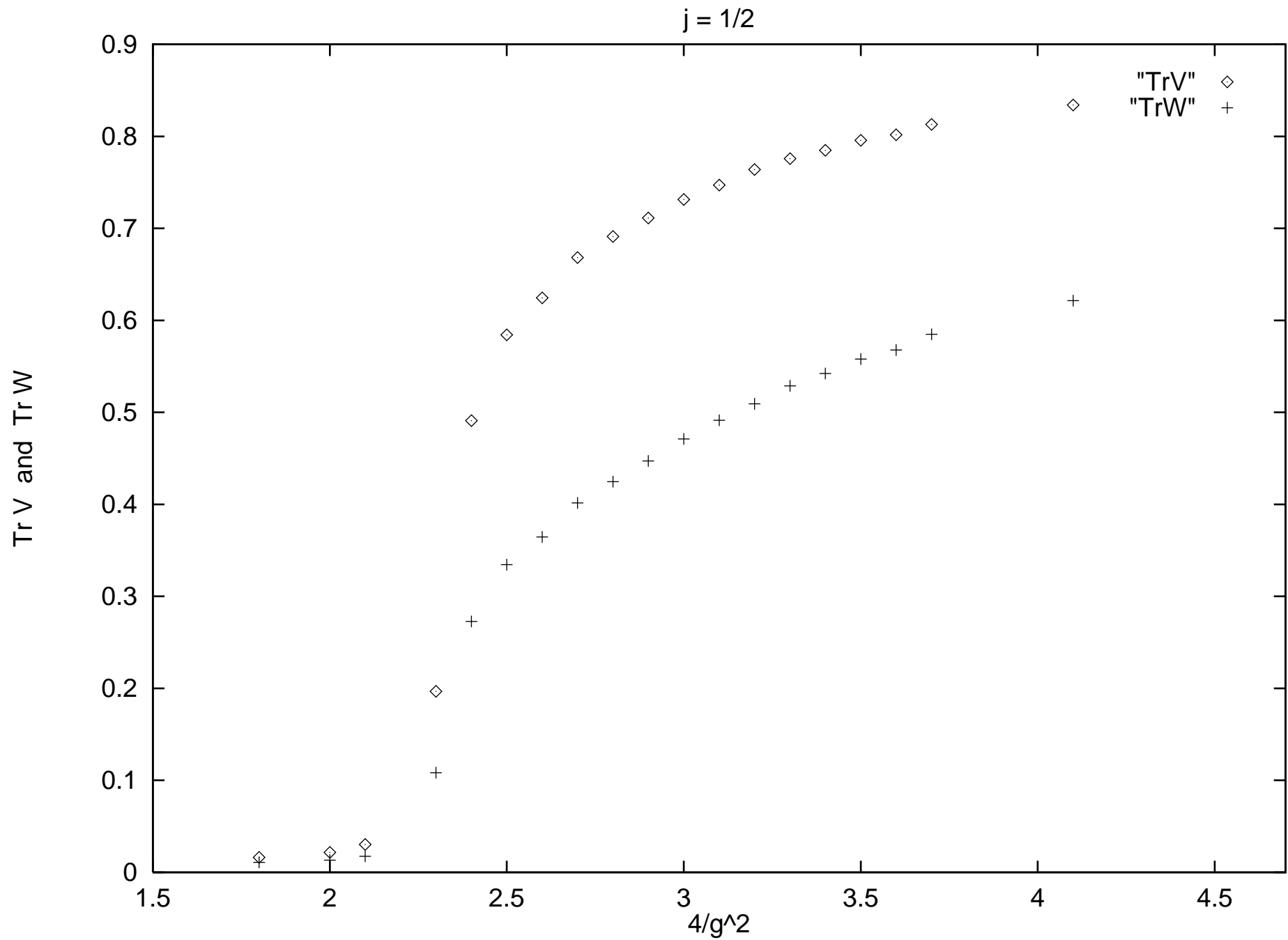
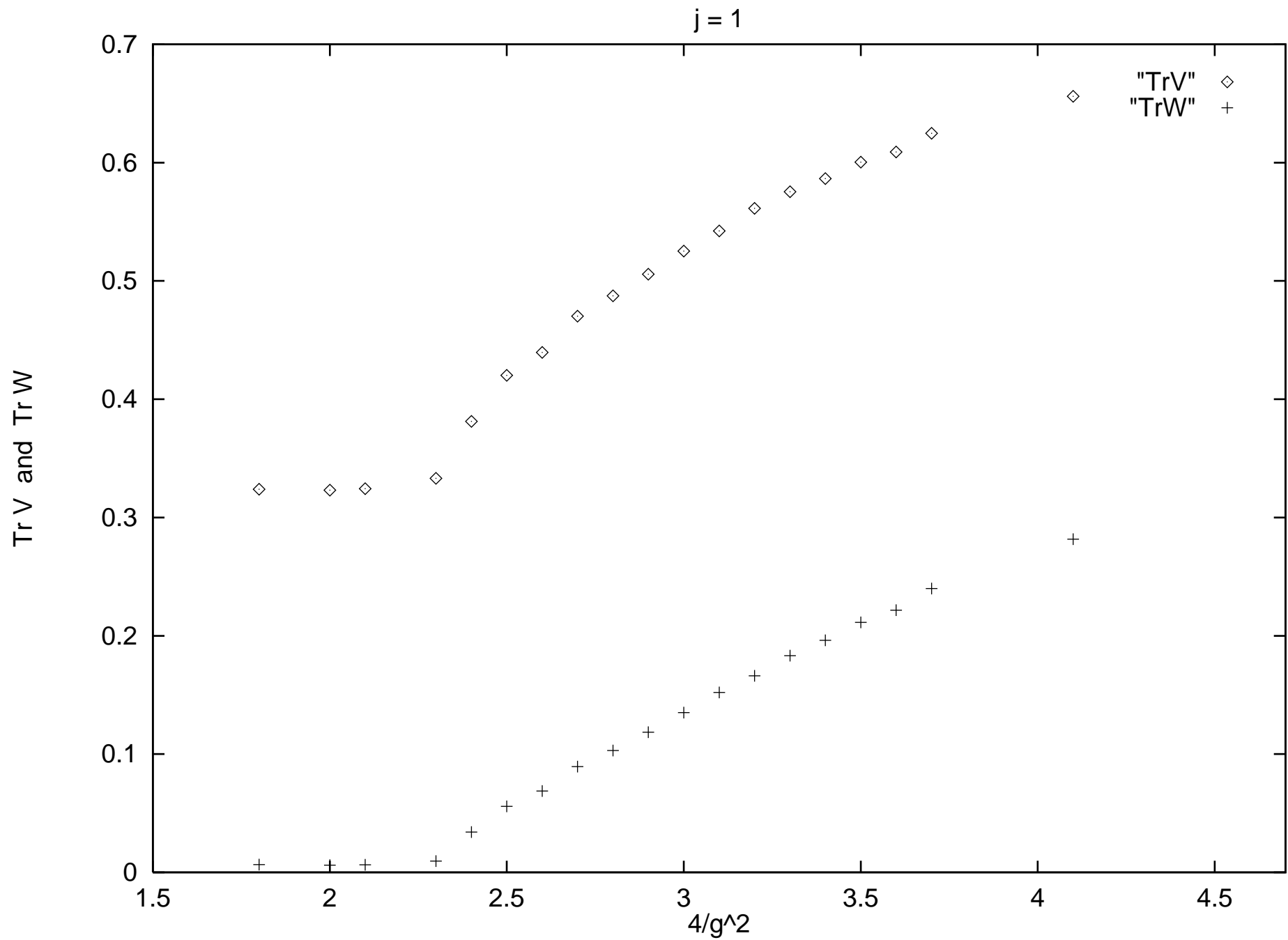
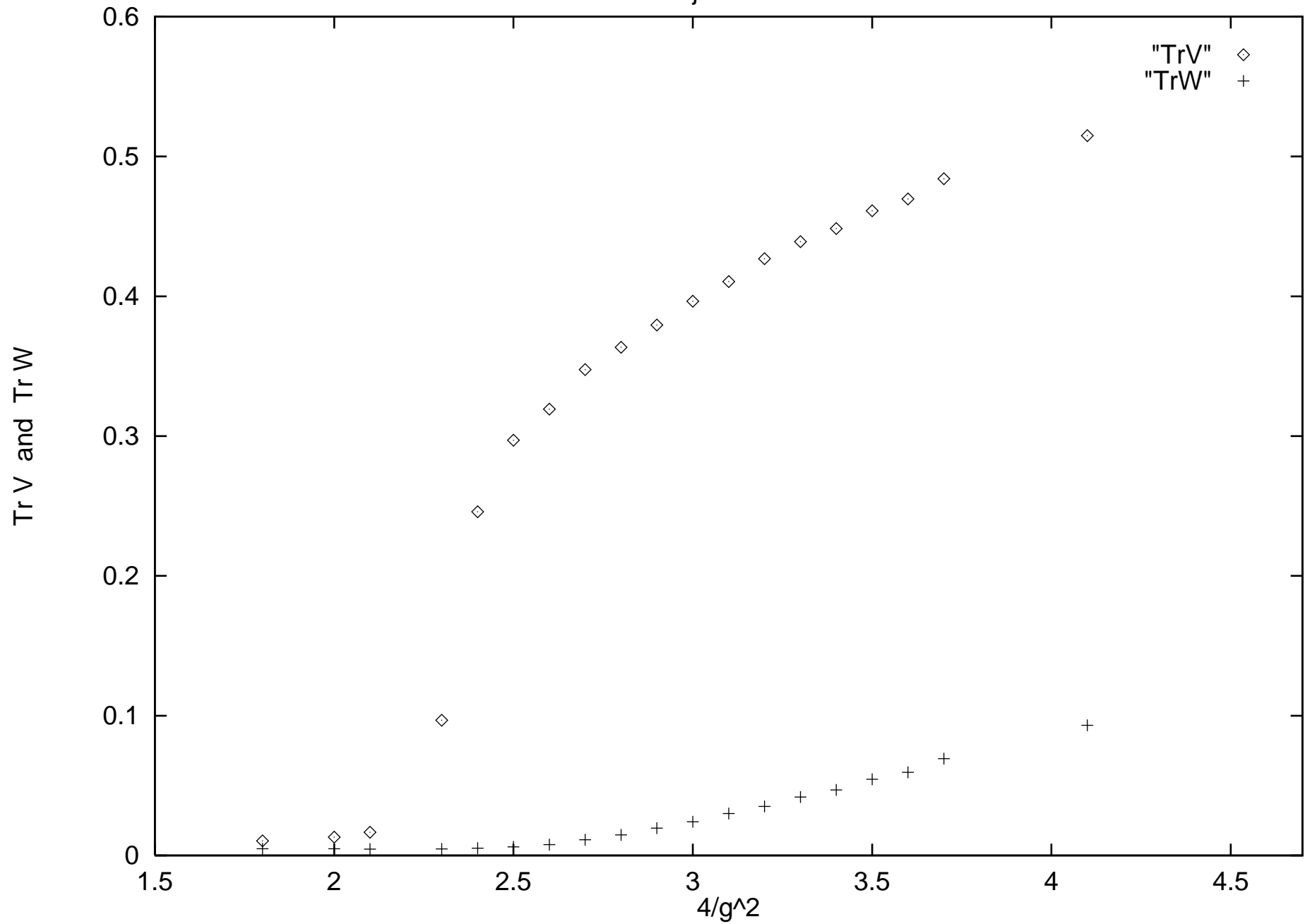
\end{document}